\documentclass[aps,twocolumn]{revtex4}
\usepackage{graphicx,subfigure}
\usepackage{amsmath,amssymb}
\usepackage{color}

\newcommand{\be}{\begin{eqnarray}}
\newcommand{\ee}{\end{eqnarray}}

\newcommand{\pa}{\partial}

\newcommand{\la}{\lambda}
\newcommand{\om}{\omega}

\newcommand{\rar}{\rightarrow}

\begin{document}

\title{Quantum and thermal fluctuations  in quantum mechanics and  field theories \\ from
a new version of semiclassical theory   }

\author{M.A.~Escobar-Ruiz$^{1,3}$}
\email{mauricio.escobar@nucleares.unam.mx}

\author{E.~Shuryak$^2$}
\email{edward.shuryak@stonybrook.edu}

\author{A.V.~Turbiner$^{1,2}$}
\email{turbiner@nucleares.unam.mx, alexander.turbiner@stonybrook.edu}

\affiliation{$^1$ Instituto de Ciencias Nucleares, Universidad Nacional Aut\'onoma de M\'exico,
Apartado Postal 70-543, 04510 M\'exico, D.F., M\'exico}

\affiliation{$^2$  Department of Physics and Astronomy, Stony Brook University,
Stony Brook, NY 11794-3800, USA}

\affiliation{$^3$  School of Mathematics, University of Minnesota,
Minneapolis, MN 55455, USA}

\date{\today}

\begin{abstract}
We develop a new semiclassical approach, which starts with the density matrix given by the Euclidean time path integral with fixed coinciding endpoints, and proceed by identifying classical (minimal Euclidean action) path, to be referred to as {\it flucton}, which passes through this endpoint. Fluctuations around flucton path are included, by standard Feynman diagrams, previously developed for instantons. We calculate the Green function and evaluate
the one loop determinant both by direct diagonalization of the fluctuation equation, and also
via the trick with the Green functions. The two-loop corrections are evaluated by explicit  Feynman diagrams, and some curious cancellation of logarithmic and polylog terms is observed. The results are fully consistent with large-distance asymptotics obtained in quantum mechanics. Two classic examples -- quartic double-well and sine-Gordon potentials -- are discussed in detail, while power-like potential and quartic anharmonic oscillator are discussed in brief. Unlike other semiclassical methods, like WKB, we do not use the Schr\"{o}dinger equation, and all the steps generalize to multi-dimensional or quantum fields cases straightforwardly.
\end{abstract}

\maketitle
\section{Introduction}

Semiclassical approximations are well known tools, both in quantum mechanical and quantum field theory applications.

Quantum mechanics itself originated from Bohr-Sommerfeld quantization conditions, and
semiclassical approximations  for the wave function -- the WKB and its extensions -- has been developed already in the early days of its development,  and are since a standard part of quantum mechanics textbooks. Unfortunately, extending such methods beyond one-dimensional cases or those with separable variables proved to be difficult.

Semiclassical approximations in quantum field theory developed differently: their  starting
point is the Feynman path integrals \cite{FH_65,Feynman_SM}, which is infinitely-dimensional anyway, and thus the dimension of quantum mechanics coordinates or number of quantum fields is of secondary importance. So, their main advantage  over the semiclassical approaches based on the Schr\"{o}dinger equation (such as WKB) is that it can be used in multi-dimensional cases.
Applications of such methods range from that by Rossi and Testa \cite{RT} in Quantum Field Theory (QFT) to recent studies of protein folding \cite{Faccioli} in statistical mechanics.

Another general advantage of the latter approach is that path integrals lead to systematic
perturbative series, in the form of Feynman diagrams, with clear rules for each order. Text-book perturbative approaches for the wave functions do not have that, and basically are never used beyond say first and second orders.

Of course,  the higher level of generality comes with a heavy price. While classical part is relatively simple, already at one-loop level one needs to calculate determinants of certain differential operators. At two and more loops Feynman diagrams need to be evaluated on top of  space-time dependent backgrounds: therefore those can be done in space-time representation rather than in energy-momentum one mostly used in QFT applications. Most content of this paper is the explicit demonstration of how one can do all that, in analytic form, for two classic examples -- quartic double-well and sine-Gordon potentials.

Let us now outline briefly the history of semiclassical evaluation of the path integrals in Euclidian time. Polyakov \cite{Polyakov} used the example of symmetric double-well potential to
demonstrate the physical meaning of celebrated ``instanton" solution in the non-Abelian gauge theories (he and collaborators discovered shortly before that). For pedagogical presentation of this material, including the one-loop corrections, see \cite{Vainshtein:1981wh}. Feynman diagrams and two-loop corrections have been calculated by F.~W\"{o}hler and E.~Shuryak \cite{E. Shuryak} for the double well potential, extended to three-loops in our recent two papers \cite{Escobar-Ruiz:2015nsa,Escobar-Ruiz:2015rfa} for both the double well and sine-Gordon potentials.

All the development was focused on the phenomenon of {\em quantum tunneling} through the barrier for degenerate minima. Polyakov's instanton is the classical path, coming from one minimum of the potential to the other. The instanton amplitude, evaluated in the above mentioned papers in higher orders, are approximations for the path integral with the endpoints of the path corresponding to this arrangement, corresponding physically to a ``spectral gap", the splitting between the lowest states of opposite parity for the double well potential case.

From the theoretical point of view, the instanton amplitudes and perturbative series around them, are parts of more general construction nowadays known as {\em  trans-series} which include series at small coupling constant $g$: power-like terms $\sim g^n$, exponentially small terms $\sim e^{-\frac{const}{g^2}}$, and logarithms of the coupling multiplied by such exponents, $\sim ({\log g})^k \ e^{-\frac{const}{g^2}}$. The issue of unique definition of such series is related with the  so called {\em resurgence} theory,  which provide certain relations between series near different extrema. Specific issues related to interplay between the perturbative series for trivial $x=0$ path and instanton-antiinstanton contributions are extensively discussed e.g. in \cite{Dunne}.

Even more general question -- whether these trans-series do define uniquely the whole
function, representing the path integral dependence on its parameters -- is the central item in rigorous mathematical definition of the QFT's. Related to it is the generalized definition of the path integrals, recently discussed by Witten \cite{Witten:2010zr}.
No question, still there remain many open questions related even with finite-dimensional integrals. Furthermore, even (1+0) dimensional path integrals -- quantum mechanical examples
under consideration --  still include certain open theoretical problems which continue to attract attention of physicists (and mathematicians) today.

In this paper we move from the well-trotted path of tunneling theory into a somewhat different direction. Instead of probability to go through the barrier, we evaluate the  $probability$ to find a quantum system at a certain position $x_0$ inside classically forbidden region.
It measures a ``strength" of quantum effects, a quantum nature of the problem.
In general, this probability is given by path integral in which the endpoints of the path coincide and are fixed. We will develop a semiclassical theory for this case. The corresponding classical solutions for it we will call  $fluctons$, following the old paper of one of us \cite{Shuryak:1987tr} where it was introduced. Another early paper devoted to the subject was that by Rossi and Testa \cite{RT}.

The paper is organized as follows. In section \ref{sec_setting} the general setting of the problem is explained, and the corresponding classical solutions, the fluctons, are derived in the section \ref{sec_flucton}. The next section \ref{sec_fluctuations} treats quantum oscillations around classical path to quadratic order, resulting in defining the corresponding determinant in section \ref{sec_det} and the Green function in section \ref{sec_GF}.

Somewhat unexpectedly, we found that the quantum-mechanical
potential for fluctuations around flucton background in the double-well problem allows exact analytic solution in elementary functions. Therefore we were able to find analytic expression for the scattering phase and evaluate the determinant via standard integral over its derivative.

Alternative derivation of the determinant is described in section \ref{sec_det_GF}, in which its derivative over the coupling is related to certain Feynman diagram, which is evaluated using the (closed loop) Green function.  Agreement of those results shows consistency of the
determinant and the Green function. Since this correspondence has never been used in the instanton problem, we discuss non-trivial sum rule for the Green function following from the determinant value: as shown in Appendix B, the Green function used in our previous works has passed this test.

In section \ref{sec_higher_loops} we evaluate two-loop corrections by direct evaluation of the diagrams over the flucton, with subtracted similar ``vacuum diagrams",  fluctuations around the trivial $x(\tau)=0$ vacuum. Surprisingly, all diagrams yield analytic answers.
While the individual diagrams contain logs and polylogs, they all cancel in sum, leading to a rather simple analytic answer
\footnote{Note that in the case of two-loop diagrams on the top of the instanton solutions, no such terms are present.}
. Would this property be true in two- and higher loop contributions: it is interesting open question.
Expansion of the results obtained for large displacement $x_0$ is compared with
the known asymptotic expansion of the ground state wave function in Appendix A.

The final section \ref{sec_other_apps} contains discussion of possible applications to other problems, in quantum mechanics with several variables, statistical mechanics and quantum
field theories.

\section{General setting}
\label{sec_setting}
By definition the Feynman path integral gives the density matrix in quantum mechanics \cite{FH_65}
\be \label{denmat}
\rho(x_i,x_f,t_{tot})\ =\ \int_{x(0)=x_i}^{x(t_{tot})=x_f} Dx(t) e^{i\,S[x(t)]/\hbar} \ .
\ee

Here $S$ is the usual classical action of the problem, e.g.
\[  S \ = \ \int_0^{t_{tot}}dt\, \bigg[ \frac{m}{2}{\bigg(\frac{dx}{dt}\bigg)}^2 - V(x)    \bigg] \ ,\]
for a particle of mass $m$ in a static potential $V(x)$ provides the weight of the paths in (\ref{denmat}).
Now let us move from quantum mechanics to statistic mechanics, from quantum system to thermal system, from density matrix to probability.
Step one is to rotate time into its Euclidean version $\tau=i\,t$. Step two is to define $\tau$ on a circle with circumference $\beta=\tau_{tot}$. Such periodic time is known as the Matsubara time, and the density matrix of quantum system is related to probability for thermal system with temperature
\be
T=\hbar/\beta \ .
\ee
Periodicity of the path implies that there is only one end parameter $x_i=x_f=x_0$.
The ensemble of such paths represent equilibrium quantum statistical mechanics at temperature $T$, or, at $T\rightarrow 0$, the ground state of the quantum system.
See details of such setting in \cite{Feynman_SM} and many other sources on statistical field theory.

The main object we will be studying in this paper is the diagonal matrix elements of the density matrix in coordinate representation, giving the probability for the specific coordinate value $x_0$ (of the field $\phi_0$) to be found in this ensemble.
The basic expression for it we will use below is a path integral with endpoints fixed and coincided
\be
P(x_0,\beta) =\int_{x(0)=x_0}^{x(\beta)=x_0} Dx(\tau) e^{-S_E[x(\tau)]/\hbar} \ .
\label{P}
\ee
thus, we will consider all (closed) trajectories starting and ending at $x_0$,
where $S_E \ = \ \int_0^{\beta}d\tau\, [ \frac{m}{2}{(\frac{dx}{d\tau})}^2 + V(x)]$.
There are two basic limits of this expression (\ref{P}). One is at large $\beta$, or low $T$.
Using standard definition of the density matrix in terms of states with definite energy
\be
P(x_0,\beta)=\sum_n |\psi_n(x_0)|^2 e^{-E_n \beta} \ ,
\ee
one sees that this limit $P$ corresponds to the lowest -- the ground state
\be P(x_0,\beta\rightarrow \infty) \sim |\psi_0(x_0)|^2 \ee
In the opposite case of small $\beta$ the circle is small and one can ignore time dependence
of the paths. In this limit
\be
P(x_0,\beta) \sim e^{- \frac{V(x_0)}{T}}\ ,
\ee
corresponding to classical thermal distribution in a potential $V$. Needless to say, the
expression is correct for any $T$.

\begin{figure}[htbp]
\begin{center}
\includegraphics[width=6cm]{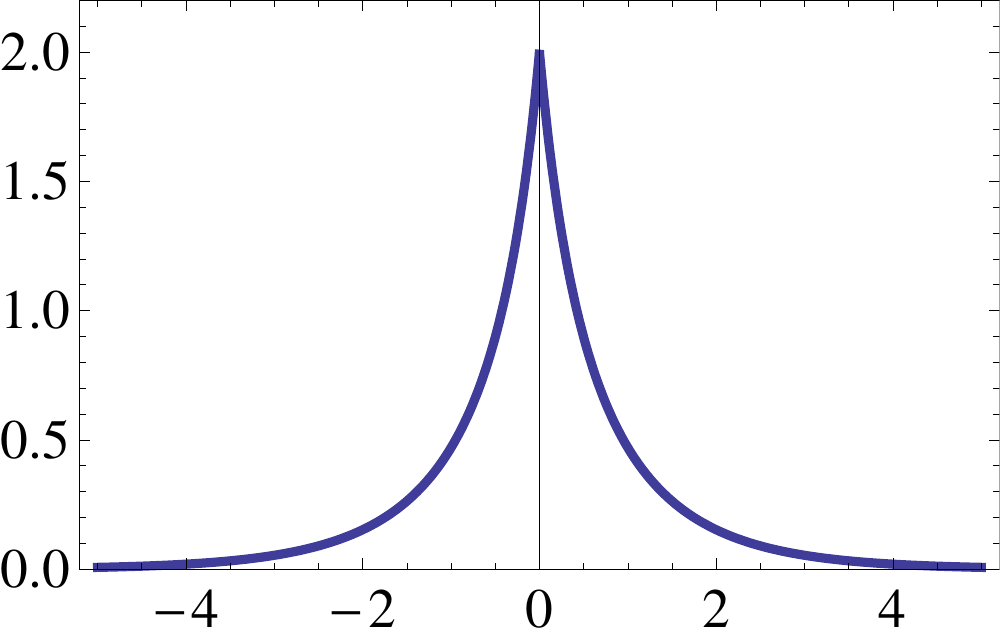}
\includegraphics[width=6cm]{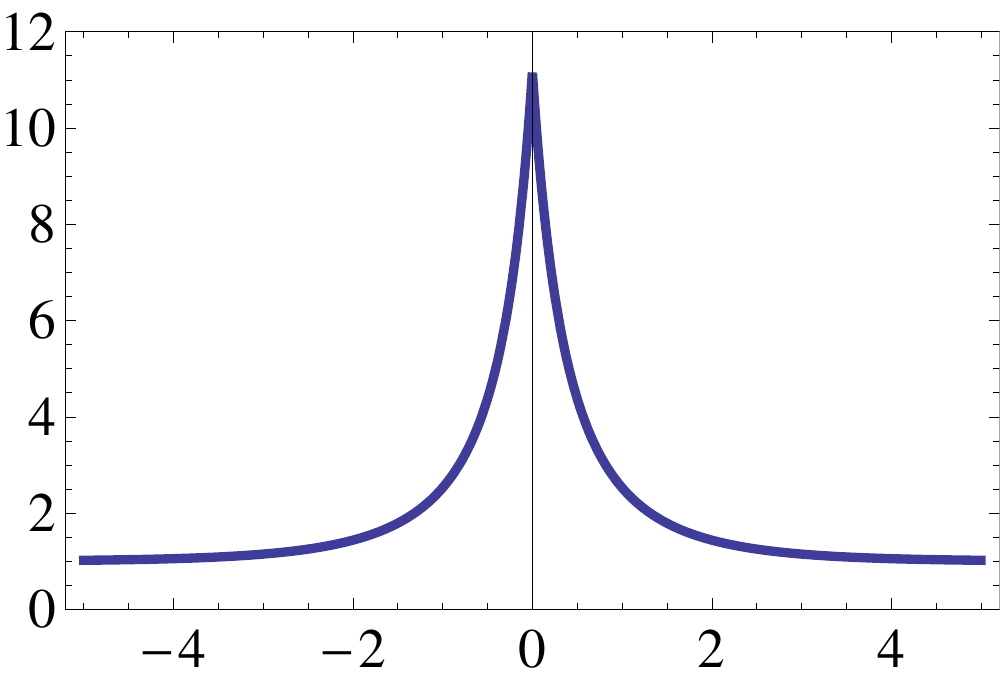}
\caption{Time dependence of the classical flucton solution $y_{fluct}(\tau)$, see (25) (upper plot) and the corresponding
potential $(1+W)$, see (37) of the fluctuations (lower plot), both for $x_0=2,\lambda =0.1$.}
\label{fig_fluct}
\end{center}
\end{figure}

\section{The classical paths: fluctons}  \label{sec_flucton}
For pedagogical reasons, we will proceed using particular examples, for which
expressions can be simple enough to allow analytic evaluation of all quantities.
The main idea is that in Euclidean time the effective potential flips and the classical
minimum becomes a maximum. Therefore classical paths  with E=0 ``slipping down" from a maximum to any point
 exist.


{\it (I).}\ It is hard not to start with the harmonic oscillator, as the first example.
One can always select units in which the particle  mass $m=1$ and the
oscillator frequency $\om=1$, so that our Lagrangian is written as
\be
  L_E = \frac{\dot{x}(\tau)^2}{2} +  \frac{x(\tau)^2}{2} \ .
\ee
Note that, for positivity, the Euclidean sign change we apply not to the kinetic, but to the potential term.
Anyway, in time $\tau$ the oscillator does not oscillate but relaxes, the classical equation of motion (EOM) produce solutions of the kind $e^\tau,e^{-\tau}$. The flucton solution at $E=0$
on a circle with circumference $\beta$ can be easily found as their superposition
satisfying
\be
    x(0)\ =\ x(\beta)\ =\ x_0\ ,
\ee
namely,
\be
   x_{flucton}(\tau)\ =\ x_0\ \frac{\left( e^{\beta - \tau}\ +\ e^\tau \right)}{e^\beta + 1}
    \ .
\ee
defined for $\tau \in [0,\beta]$.
At low $T$ (or large $\beta$) it is convenient, due to periodicity in $\tau$, to shift its range to  $\tau \in [-\beta/2,\beta/2]$.
At zero $T=1/\beta$ the range becomes infinitely large, and solution becomes
simply $x_0 e^{-| \tau |}$. At high $T$, on the other hand,
the ``thermal circle" gets small $\beta \rightarrow 0$, it can be just approximated by $x_0$.

The  classical action of such a path is
\be
   S_{flucton}\ =\ x_0^2 \ \tanh\bigg(\frac{\beta}{2}\bigg) \ ,
\ee
it tells us that the particle distribution
\be
  P(x_0) \sim \exp\left(- \frac{x_0^2}{\coth (\frac{\beta}{2})}\right) \ ,
\ee
is Gaussian at any temperature. Note furthermore, that the width of the distribution
\be
 <x^2>\ =\ \frac{1}{2} \coth \bigg(\frac{\beta}{2}\bigg)\ =
     \ \frac{1}{2} + \frac{1}{e^\beta -1}\ ,
\ee
can be recognized as the ground state energy plus one due to thermal excitation. These results are, of course, very well known, see e.g. Feynman's Statistical Mechanics \cite{Feynman_SM}.

{\it (II).}\  Our next example is the symmetric power-like potential
\be
\label{V2N}
   V\ =\ \frac{g^2}{2}\  x^{2N}\ ,\quad N=1,2,3,\ldots \ ,
\ee
for which we discuss only the zero temperature $\beta=1/T \rightarrow \infty$ case.
The (Euclidean) classical equation at zero energy $\frac{\dot{x}^2}{2} = V(x)$ has the following solution
\be
   x_{fluct}(\tau) =  \frac{x_0}{\left(1\ +\ g (N - 1) x_0^{N - 1} |\tau| \right)^{N - 1}}
   \ ,\ x_0>0,
\label{eqn_x2N}
\ee
with the action
\be
     S[x_{fluct}]\ =\   \frac{2\,g\,x_0^{N+1}}{N+1} \ ,
\ee
hence
\be
\label{Px_o}
    P(x_0) \sim \exp \left(-\frac{2\,g\,x_0^{N+1}}{N+1} \right) \ ,
\ee
which is in a complete agreement with WKB asymptotics at $x_0 \rar \infty$\
\footnote{It is worth noting that $\sqrt{P(x_0)}$ in (\ref{Px_o}) taken as parameterless variational trial function provides very accurate ground state energy for the potential (\ref{V2N}), see \cite{Turbiner:1979}}.

{\it (III).}\ The third example is the anharmonic oscillator potential
\be
\label{AHpot}
     V\ =\ \frac{1}{2}  x^{2}\ (1+g\,x^2)\ , \qquad g>0\ ,
\ee
at zero temperature $\beta=1/T \rightarrow \infty$. The classical flucton solution with the energy $E=0$ is given by
\be
     x_{fluct}(\tau) \ = \ \frac{\sqrt{g}\,x_0}{ \cosh(\tau) +\sqrt{1+g\,x_0^2}\,\sinh(\tau)} \ ,
\ee
which leads to the flucton action
\be
\label{AHaction}
   S[x_0]\ =\  \frac{2}{3}\,\frac{  {(1+g\,x_0^2)}^{\frac{3}{2}} -1}{g}   \ .
\ee
In the limit $g\rightarrow 0$ we recover the action of the harmonic oscillator and at $x_0\rightarrow \infty$ we obtain
\be
  S[x_{fluct}(\tau)] \ = \   \frac{2 \sqrt{g}}{3}x_0^3+\frac{1}{\sqrt{g}}x_0-\frac{2}{3 g} + O(\frac{1}{x_0}) \ .
\ee
in complete agreement with the asymptotic expansion of the ground state wave function squared (see Appendix A)\ \footnote{It is worth noting that $\exp{(-S[x_0]/2)}$, see (\ref{AHaction}), taken as parameterless, $g$-dependent, variational trial function provides very accurate ground state energy for the potential (\ref{AHpot}) for any value of $g \geq 0$, see \cite{Turbiner:1979}}.

However, for the most detailed studies we select two other examples.

{\it (IV).}\  One is the {\em quartic one-dimensional potential}
\begin{equation}
   V(x) \ = \  \la\,{(x^2-\eta^2)}^2 \ ,
\end{equation}
with two degenerate minima. Tunneling between them is described by well known instanton solution
\be
 x_{inst}(\tau) \, = \, \eta\,\tanh(\frac{1}{2}\om(\tau-\tau_c))\ ,
\label{instanton}
\ee
assuming that $\om^2=8\,\la\,\eta^2$. Note that the instanton has arbitrary time location $\tau_c$, while the flucton does not.

We will discuss both the {\em weak coupling } limit of small $\la$, and the {\em strong coupling} limit of large $\la$. In fact, the transition between them  happens when
the instanton action $S[x_{inst}(\tau)]=1/12\lambda$ is larger or smaller than one, respectively.

Standard steps are selecting units for $\eta$ such that $\om=1$ and shifting the coordinate by it,
\be
 x(\tau) = y(\tau) + \eta \ ,
\ee
so that the potential (21) takes the form
\be
 V = \frac{y(\tau)^2}{2} \left(1 + \sqrt{2\la} y(\tau) \right)^2 \ ,
\ee
corresponding to harmonic oscillator well at small $y$.

The {\em flucton} solution,   the minimal action path for the path integral (\ref{P}), in which the path is forced to pass through the point $x_0$ at $\tau=0$
%
now takes the form
\be
 y_{fluct}(\tau) = \frac{x_0}
 {e^{|\tau |}(1 + \sqrt{2\la}\ x_0 ) - \sqrt{2\la}\ x_0 } \ ,
\label{eqn_fluct}
\ee
We remind that  in  zero $T$ case, or infinite circle $\beta\rightarrow \infty$, $\tau \in (-\infty,\infty)$,
and solution exponentially decreases to both infinities, see Fig.\ref{fig_fluct}.
Its generalization to finite $T$ is straightforward.

The action of this solution is
\be
  S[y_{fluct}]= x_0^2 (1 + \frac{2 \sqrt{2\la} x_0}{3})\ ,
\ee
and thus in the leading semiclassical approximation the probability to find the particle at $x_0$ takes the form
\be
  P(x_0)\sim \exp\left(-x_0^2  - \frac{2\sqrt{2\la}}{3} x_0^3\right)
\ee
In the weak coupling limit only the first term remains, corresponding to Gaussian ground state wave function of the harmonic oscillator. In the strong coupling limit the second term is dominant, and the distribution then corresponds to well known cubic dependence on the coordinate. These classical-order results are of course the same as one gets from a standard WKB approximation.

When $|x_0|<\eta$ the classical flucton solution can be constructed from the pieces of the instanton and antiinstanton solutions. In this region, fluctons, instantons and antiinstantons are distinct classical paths, all contributing to the path integral (\ref{P}).
\bigskip

{\it (V).}\  Our last example is the {\em sine-Gordon potential}
\begin{equation}
    V \ = \  \frac{1}{g^2}(1-\cos(g\,x)) \ ,
\end{equation}
with infinite number of degenerate vacua. Tunneling between adjacent vacua is described by well known instanton solution
\be
    x_{inst}(\tau) \ = \ \frac{4}{g} \arctan(e^{\tau})  \ .
\label{instantonSG}
\ee
In the zero temperature case, or very large circle $\beta\rightarrow \infty$, the flucton
solution has a very simple form
\be
  x_{fluct}(\tau)\ =\ \frac{4\,\text{arccot}\bigg[ e^\tau\,\cot(\frac{g\,x_0}{4}) \bigg]}{g}
  \ .
\label{eqn_fluctSG}
\ee
The classical action for this solution is
\be
 S[x_{fluct}]= \frac{16\,\sin^2(\frac{g\,x_0}{4})}{g^2} \ ,
\ee
and, thus, in the leading semiclassical approximation the probability to find the particle at $x_0$ takes the form
\be
    P(x_0)\sim \exp[-\frac{16\,\sin^2(\frac{g\,x_0}{4})}{g^2}] \ .
\ee

\section{Fluctuations around the classical path} \label{sec_fluctuations}
Now we turn to quantum fluctuations around the classical path
\be
      y(\tau) = y_{fluct}(\tau)+f(\tau)\ ,
\ee
which is absent in the instanton case. Let us put this expression into the action and expand
it to the needed order in $f$. But before we do so, let us remind the reader
that, by the definition, all paths should pass through the same point at $\tau=0$ and, thus,
there is an important condition
\be
        f(0)=0  \ .
\label{eqn_cond}
\ee

Since the classical path is a local minimum of the action, therefore there is no term $O(f^1)$. Small fluctuations are described by the Lagrangian
\be
  L={ \dot{f}(\tau )^2\over 2}+ V''(y_{fluct}) \frac{f(\tau)^2}{2}\ +\ O(f^3)\ ,
\ee
where we used a short hand notations $V''(y_{fluct})={\pa^2 V(y)/{\pa y}^2 }|_{y=y_{fluct}}$. Its variation leads to Schr\"{o}dinger-like equation with the potential $V''$.

For harmonic oscillator this potential  $V''$ is just a constant,
so in this case the fluctuations do not depend on the classical path.
Higher order derivatives of $V$ all vanish, hence, in this case all
fluctuations are just Gaussian.

For quartic double-well potential for the famous classical solution $x_{inst}(t)$ (\ref{instanton}), the $instanton$, the potential entering has the well known form
\be
 V''(y_{fluct}) = \om^2 \left(1 - \frac{3}{2 \cosh^2(\om \tau/2)} \right)  \ .
\label{inst_V"}
\ee
This potential is one of few exactly solvable quantum mechanical problems. There are two bound states, the famous zero mode with eigenvalue zero and another state with eigenvalue $\frac{3}{4 \om^2}$, as well as the continuum of unbound states with eigenvalue above $\om^2$. Since one has the analytic expression for the scattering phase $\delta_p$,  the determinant has been evaluated so to say ``by definition", using complete set of states, for a review see e.g. \cite{Vainshtein:1981wh}. A new relation between the determinant and the Green function for the instanton  we will discuss in section \ref{sec_sum_rule}.

In the case of flucton classical solution (\ref{eqn_fluct}) the potential of the fluctuations we put into the form
\[
 V''(y_{fluct})  =1 + W \ ,
\]
where
\be
  W=\frac{6 X (1 + X) e^{|\tau |}}{(e^{| \tau |} -X  + X e^{| \tau |})^2}\ .
\label{fluct_V"}
\ee
 The classical path depends on 3 parameters of the problem, $\la, x_0$ and $\om$ (which we already put to 1): but in $W$ the first two appear in one combination only
\be
 X\ \equiv \ x_0 \,\sqrt{2\,\la}\ .
\label{eqn_X}
\ee
This observation will be important in Section \ref{sec_det}. An example of $(1+W)$ is shown in Fig.\ref{fig_fluct}(lower plot). Note that $W$ exponentially decreases at large $\tau$.

In the sine-Gordon case the potential of the fluctuations has the following form
\[
   V''(y_{fluct}) = \ \frac{1}{{(1+e^{-2\,\tau}\tan^2\,({\tilde X}))}^2   }
\]
\be
    [1+e^{-4\,\tau}\tan^4\,({\tilde X}) - 6\,e^{-2\,\tau}\tan^2\,({\tilde X})]\ ,\ \tau>0\ ,
\label{fluct_V"SG}
\ee
where the relevant combination of parameters is
\be
{\tilde X}\ \equiv \ \frac{g\,x_0}{4} \ ,
\label{eqn_Xt}
\ee
cf (\ref{eqn_X}).

\section{The flucton determinant}
\label{sec_det}

The operator governing quadratic fluctuations around flucton is
\be
 O f\ \equiv\ -\ddot{f}(\tau)+ 
  V''(y_{fluct})  f(\tau) \ ,
\label{eqn_operator}
\ee
where the derivative has already been described above (\ref{fluct_V"}).
At large $|\tau |$ the nontrivial part of the potential disappears and solutions have a generic form
\be
\psi_p(\tau) \sim \sin(p\,\tau +\delta_p ) \ ,
\ee
where for momentum $p$, only the scattering phase $\delta_p$ depends on the potential. The eigenvalues of the operator $O$ are, for the double well example (\ref{fluct_V"}),
simply,
\be
   \la_p=1+p^2 \ ,
\ee
and the determinant $\text{Det}\, O$ is their infinite product. Its logarithm is the sum
\be
    \log \,\text{Det}\, O\ =\ \sum_n \log(1+p_n^2) \ ,
\ee
where the sum is taken over all states satisfying zero boundary condition
on the boundary of some large box.

Taking the path integral over fluctuations around the classical path, in the Gaussian approximation, leads to the following standard expression
\[
   P(x_0) = \frac{ \exp \left( -S[x_{flucton}]\right)}{\sqrt{\text{Det}\,(O_{flucton})} }
\]
\be
\label{Px0}
   \times \left[1+O(two \,\,and \,\, more\,\,  loops) \right]\ ,
\ee
with $O_{flucton} = O$ defined in (41).
In this section we discuss numerical evaluation of the determinant:  another method will be discussed in the section \ref{sec_det}, after we will derive the corresponding Green function for the fluctuations in section \ref{sec_GF}. Calculation of two and more loop corrections via Feynman diagrams will be discussed in section \ref{sec_higher_loops}.



As it is well known, the nontrivial part of the problem is not in the eigenvalues themselves, but in the counting of levels.  Standard method (see e.g. $\S$ 77 of \cite{LLstatmech})
vanishing boundary conditions at the boundary of some large box, at $\tau=L$, leads to
\be
  p_n L +\delta_{p_n} = \pi\, n \ ,\ n=1,2,\ldots \ .
\ee
At large $L$ and $n$ one can replace summation to an integral, resulting in the generic expression
\[
 \log \text{Det}\, O\ =\ \sum_n \log(1+p_n^2)\
\]
\be
 =\ \int_0^\infty \frac{dp}{\pi}  \frac{d\delta_p}{dp}  \log(1+p^2) \ .
\label{eqn_logdet}
\ee

After using few different numerical methods for particular values of the parameter $X$, we discovered that there exist {\it exact} (non-normalized) analytic solution for the eigenfunctions of the form
\be
   \psi_p(\tau)\ \sim\  \sin\left( p\,\tau + \Delta(p,\tau) \right)\ F(p,\tau) \ ,
\label{eqn_sol}
\ee
with the following two functions
\[
 \Delta(p,\tau)= \text{arctan}\,\bigg[\frac{ -3 p\,(1 + 2 X)}{1 - 2 p^2 + 6 X + 6 X^2}\bigg]
\]
\[
 +\ \text{arctan} \bigg[\frac{N}{D}\bigg] \ ,
\]
where

\[
  N\ =\ 3 p [1 + 2  X + X^2 - X^2 e^{-2\tau}] \ ,
\]

\begin{figure}[h]
\begin{center}
\includegraphics[width=3.0in,angle=0]{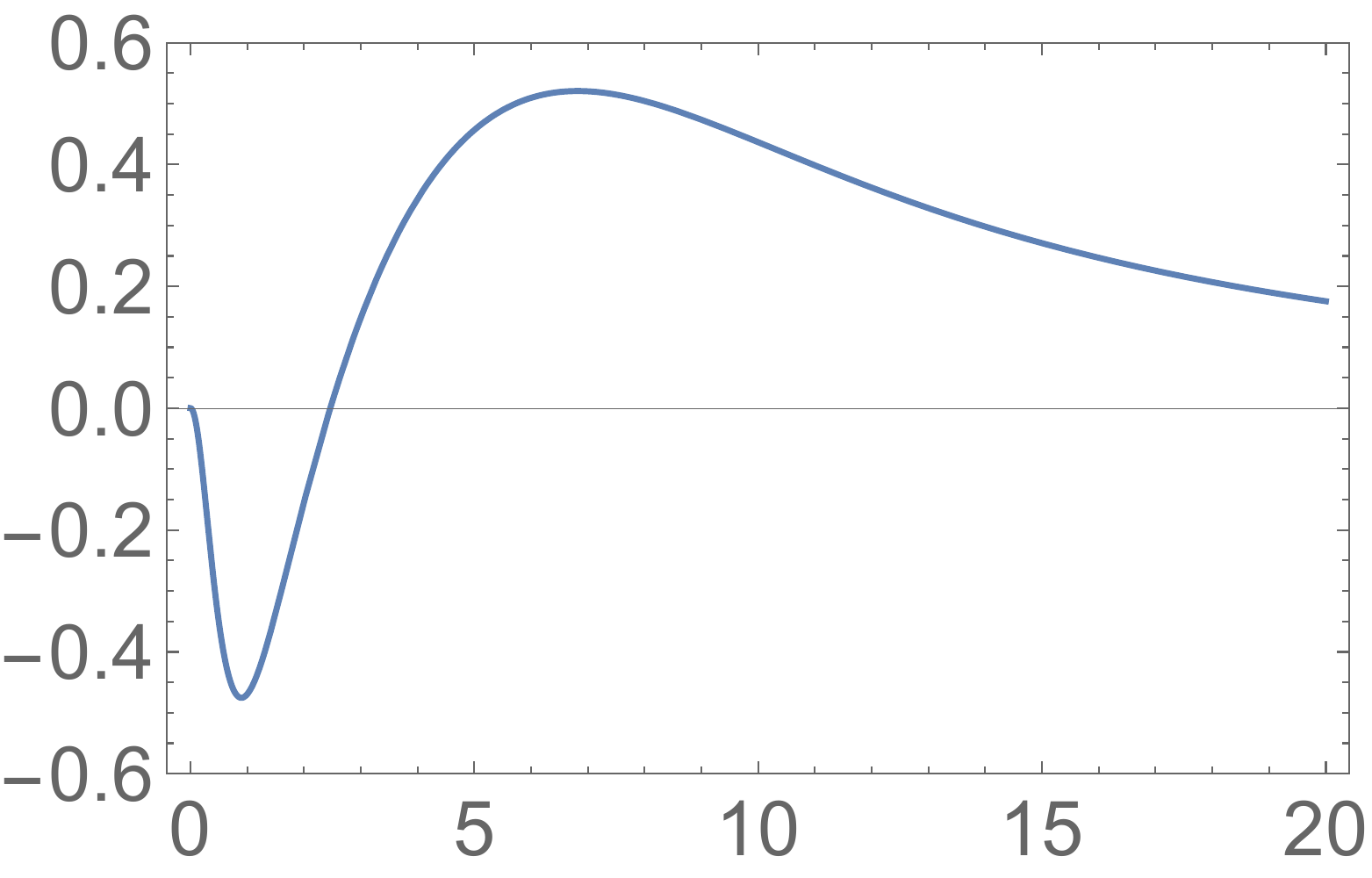}
\includegraphics[width=3.0in,angle=0]{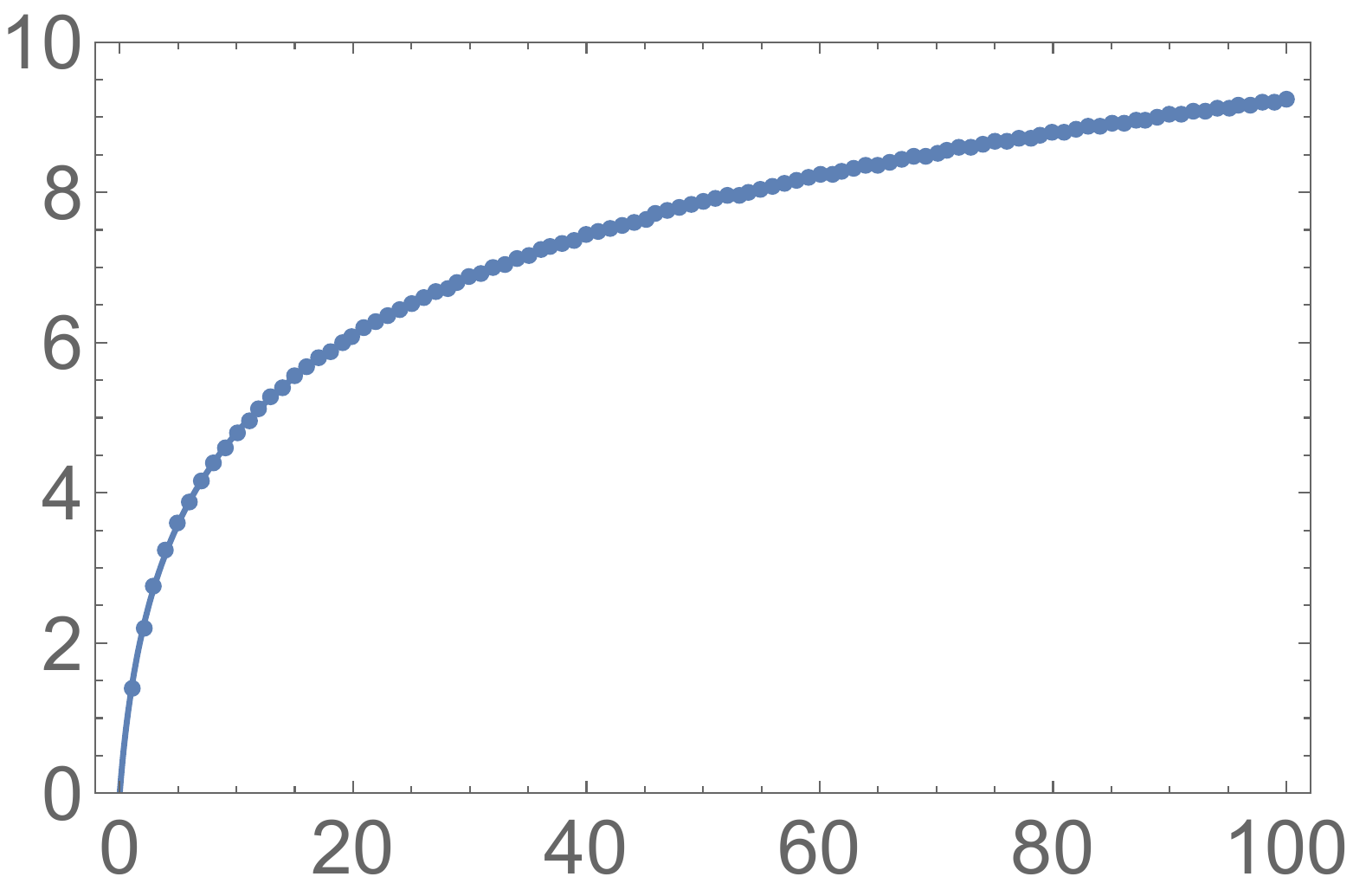}
\caption{(a)The integrand of  (\ref{eqn_logdet}), $\log(1+p^2)d\delta_p/dp$, versus $p$, for $X=4$.
(b) The integral  (\ref{eqn_logdet}) vs parameter $X$: points are numerical evaluation, line is defined in the text.
}
\label{fig_ddeltadp}
\end{center}
\end{figure}

\begin{widetext}
\[
  D\ =\  (2 p^2-1)(1+X^2) -
  2 e^{-\tau} \big(2 (1 + p^2) -
  e^{-\tau} (2 p^2-1)\big) X + (2 p^2-1)e^{-2\tau} -
     4 e^{-\tau} (1 + p^2) \ ,
\]

$$
  F(p,\tau)\ =\ \frac{1}{(e^{\tau} -X + e^{\tau} X)^2}
 \times
\bigg[e^{4 \tau} (1 + 5 p^2 + 4 p^4) +
     4 e^{3 \tau} (1 + p^2) \bigg(2 - 4 p^2 + e^{\tau} (1 + 4 p^2)\bigg) X + $$ $$
     6 e^{2\tau} \bigg(3 + p^2 + 4 p^4 + 4 e^{\tau} (1 - p^2 - 2 p^4) + 
        e^{2\tau} (1 + 5 p^2 + 4 p^4)\bigg) X^2 +
     4 e^{\tau} \bigg(2 (1 - p^2 - 2 p^4) + $$ $$
        6 e^{2\tau} (1 - p^2 - 2 p^4) + 3 e^{\tau} (3 + p^2 + 4 p^4) + 
        e^{3 \tau} (1 + 5 p^2 + 4 p^4)\bigg) X^3 + \bigg(1 + 5 p^2 + 4 p^4 + $$ $$
        8 e^{\tau} (1 - p^2 - 2 p^4) + 8 e^{3 \tau} (1 - p^2 - 2 p^4) +
        6 e^{2\tau} (3 + p^2 + 4 p^4) +
        e^{4 \tau} (1 + 5 p^2 + 4 p^4)\bigg) X^4\bigg]^{1/2} \ .
$$

It is important that at $\tau=0$ the solution (\ref{eqn_sol}) goes to zero:  according to the flucton definition, all fluctuations at this point must vanish (\ref{eqn_cond}).
It is the condition which fixes the scattering phase.

At large time, where all terms with decreasing exponents in $\Delta(p,\tau)$ disappear and
the remaining constant terms define the scattering phase, we need
\be \label{eqn_delta}
  \delta_p = \arctan \bigg[\frac{3 p(1 + 2 X)}{1 - 2 p^2 + 6 X + 6 X^2}\bigg]
     - \arctan\bigg[\frac{3p}{1 - 2p^2}\bigg] \ .
\ee
\end{widetext}
Comments: \\[5pt]
(i) the scattering phase is $O(p)$ at small $p$; \\
(ii) it is $O(1/p)$ at large $p$ and, thus, there must be a maximum at some $p$; \\
(iii) for $X=0$ two terms in (\ref{eqn_delta})  cancel out. This needs to be the case since in this limit the
nontrivial potential $W$ of the operator also disappears;\\
(iv) at large time the amplitude $F$ (\ref{eqn_sol}) goes to a constant, as it should.\\
The $\arctan$-function provides an angle, defined modulo the period, and thus it experiences jumps by $\pi$.
Fortunately, its derivative $d\delta_p/dp$ entering the determinant (\ref{eqn_logdet}) is single-valued and smooth.
The momentum dependence of the integrand of this expression for $X=4$ is shown in Fig.\ref{fig_ddeltadp}(a).
Analytic evaluation of the integral  (\ref{eqn_logdet}) was not successful, the results of the numerical evaluation are shown by points in Fig.\ref{fig_ddeltadp}(b). However, the  {\it guess} $2 \log(1+X)$, shown by the curve
in Fig.\ref{fig_ddeltadp}(b)  happens to be accurate to numerical accuracy, and thus it must be correct. We will demonstrate that it is exact below.

Since the calculation above includes only a half of the time line, $\tau>0$, and the other half is symmetric, the complete result for the $\log Det\, O$ should be doubled.  Substituting  (\ref{eqn_delta}) to  (\ref{eqn_logdet}) we obtain a (surprisingly simple) exact result
\be
   Det\,(O)\ =\ (1+X)^4 \ .
\label{eqn_Det}
\ee
Note that at $X=0$ we return to harmonic oscillator case.

\section{The Green function of the fluctuations around the flucton solution
\label{sec_GF}}

General procedure for  inversion of the operator (\ref{eqn_operator}), leading to a Green function, is different for the instanton and flucton cases.
In the instanton case the inversion is only possible in the subspace normal to the zero mode, leading to specific difficulties.
The flucton problem we discuss now has no shift symmetry (no translation invariance) and thus no zero modes. Needless to say that this symmetry is killed by the boundary condition at the fixed moment, $f(\tau=0)=0$.

The corresponding equation to be solved thus is
\be
  -\ \frac{\pa^2 G(\tau_1,\tau_2)}{\pa \tau_1^2}+  V'' (y_{fluct}(\tau_1)) G (\tau_1,\tau_2)
\ee
\[
 =\ \delta(\tau_1-\tau_2)  \ .
\]

Homogeneous equation (with zero r.h.s.) has two solutions

\be
f_0(\tau)\ =\  \frac{e^{\tau}}{{(e^{\tau}(1+X)-X)}^2}  \ ,
\ee
and
\begin{widetext}
\be
f_1(\tau)\ =\   \frac{e^{-\tau}}{2\,{(X-e^{\tau}\,(1+X))}^2} \bigg(8\,X^3\,(1+X)\,e^{\tau} +   12\,X^2\,{(1+X)}^2\,\tau\,e^{2\,\tau}
-   8\,X\,{(1+X)}^3\,e^{3\,\tau} + {(1+X)}^4\,e^{4\,\tau} - X^4  \bigg)
\ee

(Hereafter we only discuss half line $\tau>0$). The first solution -- would be zero mode if shift be allowed -- is exponentially decreasing at large time, the second one is increasing in time. Standard construction immediately yields the following Green function


\[
    G(\tau_1,\,\tau_2)\ =\ \frac{e^{-|\tau_1-\tau_2|}}{2\ {\big(e^{\tau_1}(1+X)-X\big)}^2\,{\big(e^{\tau_2}(1+X)-X\big)}^2}
    \bigg[ 8\ e^{\frac{1}{2}(\tau_1+\tau_2+3\,|\tau_1-\tau_2|)}\,X^3\,(1+X)
\]
\[
    -\ 8\ e^{\frac{1}{2}(3\tau_1+3\tau_2+|\tau_1-\tau_2|)}\,X\,(1+X)^3
    +\ e^{2\,(\tau_1+\tau_2)}\,{(1+X)}^4      -6\,e^{(\tau_1+\tau_2+|\tau_1-\tau_2|)}\,X^2\,{(1+X)}^2\, |\tau_1-\tau_2|
\]
\[
   +\  e^{(\tau_1+\tau_2+|\tau_1-\tau_2|)}\,\bigg(\,6\,X^4\,(\tau_1+\tau_2)\, +\, 12\,X^3\,(1+\tau_1+\tau_2)
 +\ 6\,X^2\,(3 + \tau_1+\tau_2)  + 4\,X - 1 \bigg)
\]

\be
\label{GDW}
  - e^{2\,|\tau_1-\tau_2|}\,X^4 \ \bigg] \ ,
\ee


for $\tau_1,\,\tau_2>0$ .

Similarly, in the sine-Gordon problem the same standard construction yields
the following Green function

$$
 G(\tau_1,\,\tau_2)\ =\  \frac{1}{8\,\big(\cosh(\tau_1) +\cos(2\,{\tilde X})\,\sinh(\tau_1) \big)}\times
  \frac{1}{(\cosh(\tau_2) +\cos(2\,{\tilde X})\,\sinh(\tau_2))}\times $$
$$ \bigg[ 2\,(\tau_1+\tau_2 - |\tau_2-\tau_1|)\,\sin^2(2\,{\tilde X})   +\ 8\,\cos(2\,{\tilde X})\,\sinh^2(\frac{1}{2}(\tau_1+\tau_2 - |\tau_2-\tau_1|)) $$
\be   +\ (3+\cos(4\,{\tilde X}))\,\sinh(\tau_1+\tau_2 - |\tau_2-\tau_1|)   \  \bigg]\ ,
\ee
\end{widetext}
for $\tau_1,\,\tau_2>0$ .



\section{Relating the determinant and the Green function
\label{sec_det_GF} }

\begin{figure}[h]
\begin{center}
\includegraphics[width=3.0in,angle=0]{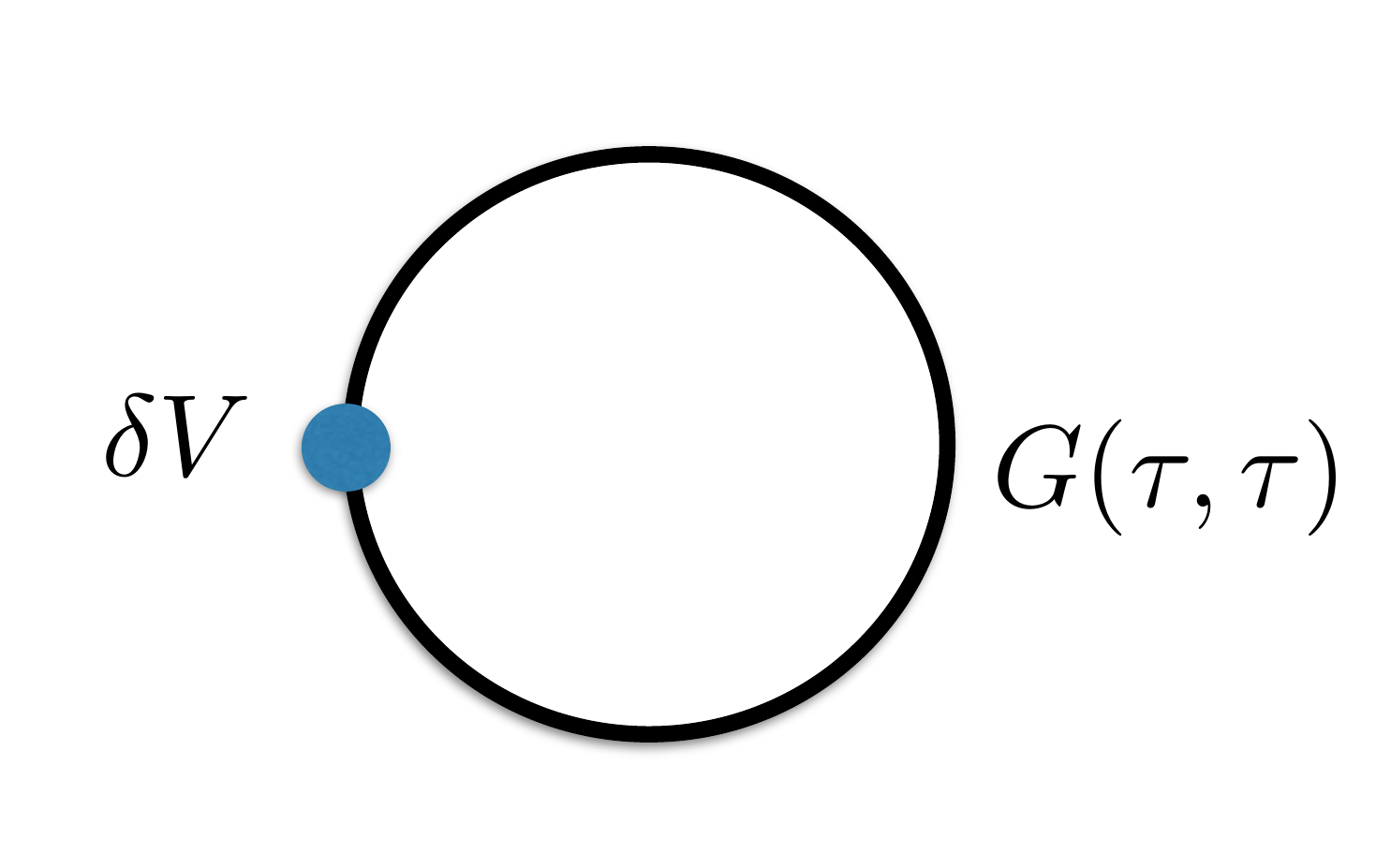}
\caption{Symbolic one-loop diagram, including variation of the fluctuation potential $\delta V$and the simplified ``single-loop"
Green function $G(\tau,\tau)$.
}
\label{fig_oneloop}
\end{center}
\end{figure}

The method we will use in this section relies on the following observation.
When the fluctuation potential depends on some parameter, it can be varied. In the case at hand (\ref{fluct_V"}), the potential we write as
\[
V_{flucton}=1+W(X,\tau) \ ,
\]
depends on the combination (\ref{eqn_X}). Its variation resulting in extra potential
\be
     \delta  V_{flucton} = \frac{\pa W}{\pa X} \delta X
\ee
is a perturbation: its effect can be evaluated by the following Feynman diagram
\be
 \frac{\pa \log \text{Det}\,(O_{flucton})}{\pa X}\ =\ \int d\tau G(\tau,\tau)
 \frac{\pa V_{flucton}(\tau)}{\pa X}\ ,
\label{relation}
\ee
containing derivative of the potential as a vertex and the ``loop",  the same point  Green function, see Fig.\ref{fig_oneloop}. This relates the determinant and the Green function
\footnote{The historical origin of this idea goes back to Brown and Creamer \cite{Brown:1978yj}, see also \cite{Corrigan:1979di}, for gauge theory instanton. Zarembo \cite{Zarembo:1995am} applied it for the monopole  and Diakonov et al, \cite{Diakonov:2004jn} for the calorons at nonzero holonomy. The reasons it has not been used for quantum mechanical instantons, and the related sum rule are discussed in Appendix   \ref{sec_sum_rule}.}
: if the r.h.s. of it can be calculated, the derivative over $X$ can be integrated back.

In the quartic double-well problem the "Green function loop" propagator is
\begin{widetext}
\be
   G(\tau,\,\tau)\ =\ \frac{1}{2 \big(X - e^\tau (1 + X)\big)^4}
\ee
$$ \times \bigg(-X^4 + 8 e^\tau X^3 (1 + X) -
   8 e^{3 \tau} X (1 + X)^3 + e^{4 \tau} (1 + X)^4 +
 e^{2 \tau} (-1 +   4 X +  18 X^2 + 12 X^3 + 12 X^2 (1 + X)^2 \tau)\bigg)\ ,$$
\end{widetext}
and the ``vertex"
\be
  \frac{\pa V_{flucton}(\tau)}{\pa X}\ =\
 \frac{6 e^\tau \big(X + e^\tau (1 + X)\big)}{\big(-X + e^\tau (1 + X)\big)^3}\ .
\ee

With these expressions  one can evaluate the r.h.s. of the relation (\ref{relation}),
and adding the same expression for negative time, one gets the result
\be
     \frac{\pa \log \text{Det}\,({O_{flucton} })}{\pa X}\ =\ \frac{4}{1+X}\ ,
\ee
which exactly agrees with the result (\ref{eqn_Det}) from the direct evaluation of the determinant using the phase shift.
So, the Green function has passed a very nontrivial test, and  we  conclude that
it is ready to be used for evaluation of two and higher loop diagrams.

In the sine-Gordon problem the corresponding simplified expression for $G(\tau,\tau)$ is
\begin{widetext}
\be
G(\tau,\,\tau) \ =\ \frac{1}{ 4\,{\bigg(1+e^{2\,\tau} +(e^{2\,\tau}-1)\cos(2\,{\tilde X})  \bigg)}^2}
   \times \bigg[ 4\,{(e^{2\,\tau}-1)}^2\cos(2\,{\tilde X}) -\cos(4\,{\tilde X})
 \, \ee
 $$
   +\ e^{4\,\tau}\big(3+\cos(4\,{\tilde X})\big)+\,8\,e^{2\,\tau}\,\tau \,
  \sin^2(2\,{\tilde X}) - 3\bigg] \ .
$$
Evaluating the one-loop diagram Fig.\ref{fig_oneloop}, we arrive at the result
\be
 \log \,\text{Det}\,(O_{flucton})\ =\ 4 \, \tan[ {\tilde X}]\ .
\ee

For the power-like potential (\ref{V2N}) the ``Green function loop" takes the form

\be
  G(\tau,\,\tau)\ =
 \frac{-1+(N-1) X_1 \tau +\left(1+X_1 (1-N )\tau \right){}^{\frac{2 N}{1-N}}}{(3 N-1) X_1}
 \ ,
\ee
where
\[
     X_1=g\,x_0^{N-1}\ ,
\]
and the ``vertex" reads
\be  {\partial V_{flucton}(\tau) \over \partial X_1}\ =  \frac{2 (2 \,N-1)\, N\, X_1}{\left((N-1) \,\tau\,  X_1-1\right){}^3}         \ ,
\ee
$(\tau<0)$. Hence, we obtain the result
\[ \log \, \text{Det} (O_{flucton})\ =\ \frac{2\,N}{N-1}\,\log X_1  \ .  \]
In the case of the anharmonic oscillator (\ref{AHpot}) the ``Green function loop" is

\be
  G(\tau,\,\tau)\ =\ \frac{\left(\sinh(\tau)+\cosh(\tau) X_2\right) }{4 X_2 \left(\cosh(\tau)+\sinh(\tau) X_2\right){}^4}
     \bigg[-6 \tau  X_2 \left(\sinh(\tau)+\cosh(\tau) X_2\right) \left(-1+X_2^2\right)
\ee
$$
  + \sinh(\tau) \bigg(4+X_2[\,\sinh(2 \tau )+3 (-3+\cosh(2\tau)) X_2
+ 3 \sinh(2\tau)X_2^2+(5+\cosh(2\tau)) X_2^3]\,\bigg)\bigg] \ ,  $$
\end{widetext}
where
\[
 X_2=\sqrt{1+g\,x_0^2}\ ,
\]
while the ``vertex" is given by
\be  {\partial V_{flucton}(\tau) \over \partial X_2}\ = \frac{12 \left(\sinh(\tau)+\cosh(\tau) \,X_2\right)}{\left(\cosh(\tau)+\sinh(\tau)\, X_2\right){}^3} \ ,
\ee
$(\tau>0)$. Thus,
\[
      \log \, \text{Det} (O_{flucton})\ =\ 2 \log[X_2 (1+X_2)] \ .
\]

\section{Higher order Feynman diagrams
\label{sec_higher_loops}}

Now, using only the tools from quantum field theory, the Feynman diagrams in the flucton background, we compute the two-loop correction to the density matrix (\ref{Px0}) for the double-well potential. In principle, the higher order diagrams are evaluated by standard rules.

Unlike  the calculations near the instanton solution \cite{E. Shuryak,Escobar-Ruiz:2015nsa},
in the case of flucton there are no zero modes and related Jacobian, so all diagrams
follow from the Lagrangian. In the quartic double-well potential, the flucton-based Green function (\ref{GDW}) was determined above and the only vertices are
triple and quartic ones
\be
  v_3(\tau)\ =\
   \frac{6 \sqrt{2\,\la} \left(X+e^\tau (1+X)\right)}{-X+e^\tau (1+X)}\ ,
\ee
\be
v_4\ =\ 24\,\la \ .
\ee
The loop corrections in (\ref{Px0}) are written in the form
\[
 \left[1+O(two \,\,and \,\, more\,\,  loops) \right] \ = \  2\,\sum_{n=0}^{\infty}\,B_n\,\lambda^n \ ,
 \quad B_0=\frac{1}{2}\ ,
\]
where $B_n=B_n(X)$. Like in the calculations near the instanton solution, we need to separate the finite flucton-related contribution for each diagram from the infinite (time-divergent) contribution without it. This is done by subtracting the same expression with ``vacuum vertices"
\be
  v_{3,0}\ =\  6 \, \sqrt{2\la} \ ,
\ee
\be
  v_{4,0}\ =\ 24\,\la \ ,
\ee
and the ``vacuum propagator"
\be
\label{GDW0}
G_0=G(\tau_1,\,\tau_2)\mid_{{}_{X\rightarrow 0}}\ =\ \frac{e^{-|\tau_1 - \tau_2|}}{2} - \frac{e^{-\tau_1 - \tau_2}}{2} \ .
\ee
(Note that (\ref{GDW0}) differs from the vacuum propagator in the instanton problem where the second term in the r.h.s. is absent. In particular, it is no longer translational invariant because of extra boundary condition at $\tau_1=\tau_2=0$ for fluctuations at the fixed point.)

The two-loop correction $B_1$ we are interested in can be written as the sum of three diagrams, see Fig.4, diagram $a$ which is a one-dimensional integral and diagrams $b_1$ and $b_2$ corresponding to two-dimensional ones.

Explicitly, we have
\begin{widetext}
\be
 a \ \equiv \ -\frac{1}{8\,\la}\,v_4\,\int_0^{\infty}[G^2(\tau,\,\tau) -
 G_0^2(\tau,\,\tau) ]d\tau  =\ \frac{3}{560 X^2 (1+X)^4}
\ee
$$   \times \bigg(24X-60X^2-520X^3 -1024 X^4 - 832 X^5 - 245 X^6
 +24 (1+X)^2 (1+2 X) (-1+6 X(1+X)) \log(1+X) $$
$$  +288 X^2 (1+X)^4 \text{PolyLog}\left[2,\frac{X}{1+X}\right]\bigg) \ , $$
here $\text{PolyLog}[n,z] =\sum _{k=1}^{\infty } z^k/k^n$ is the polylogarithm function and
\be
 b_1 \ \equiv \ \frac{1}{12\,\la}\,\int_0^{\infty}\int_0^{\infty}[v_3(\tau_1)\,v_3\,(\tau_2)G^3(\tau_1,\,\tau_2)
  -\ v_{3,0}v_{3,0}G_0^3(\tau_1,\,\tau_2) ]\,d\tau_1\,d\tau_2
\ee
$$
= \frac{1}{280 X^2 (1+X)^4} \times
  \bigg(-24X+60 X^2+520 X^3+1024X^4+832 X^5+245 X^6
 $$
$$ + 24 (1+X)^2 \left(1-4 X-18 X^2-12 X^3\right) \log(1+X) - 288 X^2 (1+X)^4 \text{PolyLog}\left[2,\frac{X}{1+X}\right]\bigg) \ ,$$

\be
b_2 \ \equiv \ \frac{1}{8\,\lambda}\,\int_0^{\infty}\int_0^{\infty}\big[v_3(\tau_1)\,v_3\,
     (\tau_2)G(\tau_1,\,\tau_1)G(\tau_1,\,\tau_2)G(\tau_2,\,\tau_2)
   -\ v_{3,0}v_{3,0}G_0(\tau_1,\,\tau_1)G_0(\tau_1,\,\tau_2)G_0(\tau_2,\,\tau_2)\big]
   \,d\tau_1\,d\tau_2
\ee
$$
  = \ -\frac{1}{560 X^2 (1+X)^4} \times
   \bigg(24X-60 X^2+1720 X^3+5136 X^4+4768 X^5+1435 X^6
   $$
$$   +24 (1+X)^2 \left(-1+4 X+18 X^2+12 X^3\right) \log(1+X) +288 X^2 (1+X)^4 \text{PolyLog}\left[2,\frac{X}{1+X}\right]\bigg) \ . $$
\end{widetext}

\begin{figure}[h!]
\begin{center}
\includegraphics[width=3.0in,angle=0]{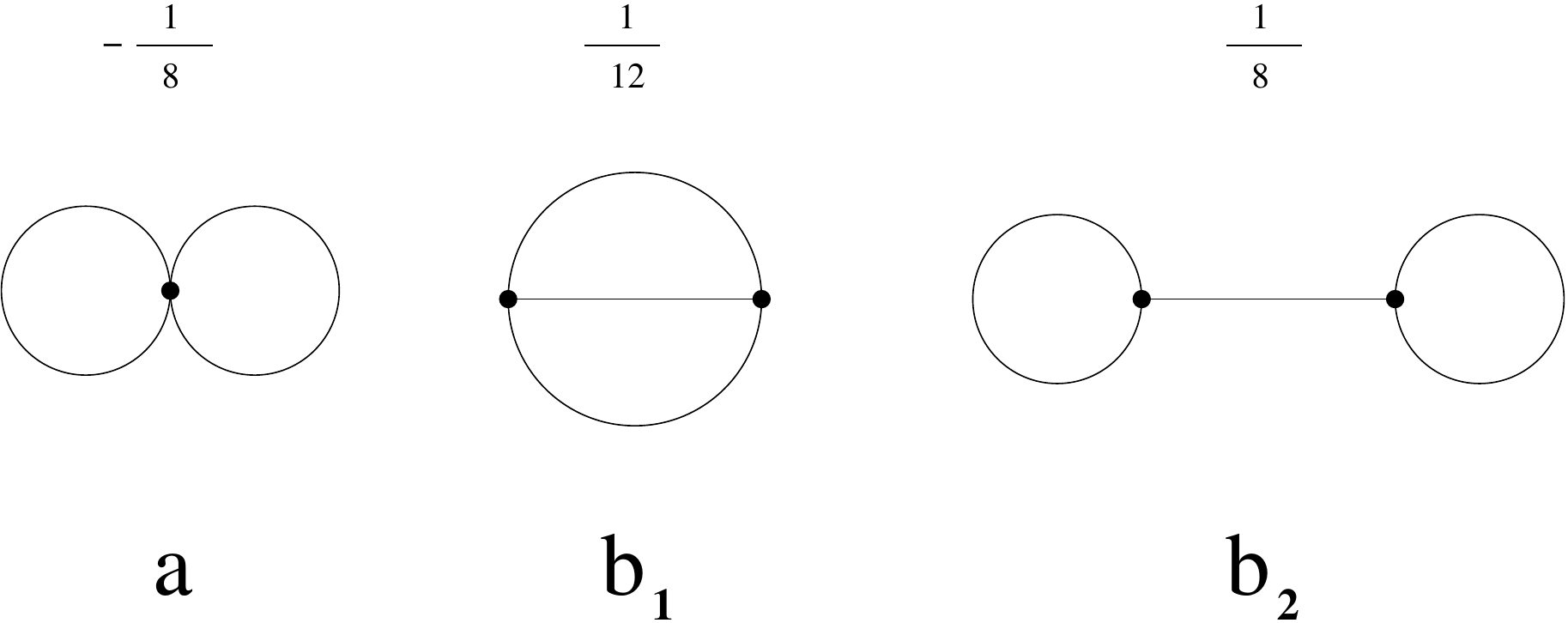}
\caption{Diagrams contributing to the two-loop correction $B_1 = a + b_1 + b_2$. The signs of contributions and symmetry factors are indicated. }
\label{2loopplot}
\end{center}
\end{figure}

Eventually, the two-loop correction takes an amazingly simple form,

\be
\label{B1}
   B_1 \, \equiv \ a+b_1+b_2 \ = \ -\frac{X (4+3 X) }{(1+X)^2} \ ,
\ee
all $\log$ and $\text{PolyLog}$ terms disappear! The results of calculations are shown on Fig.5.

\begin{figure}[b]
\begin{center}
\includegraphics[width=3.75in,angle=0]{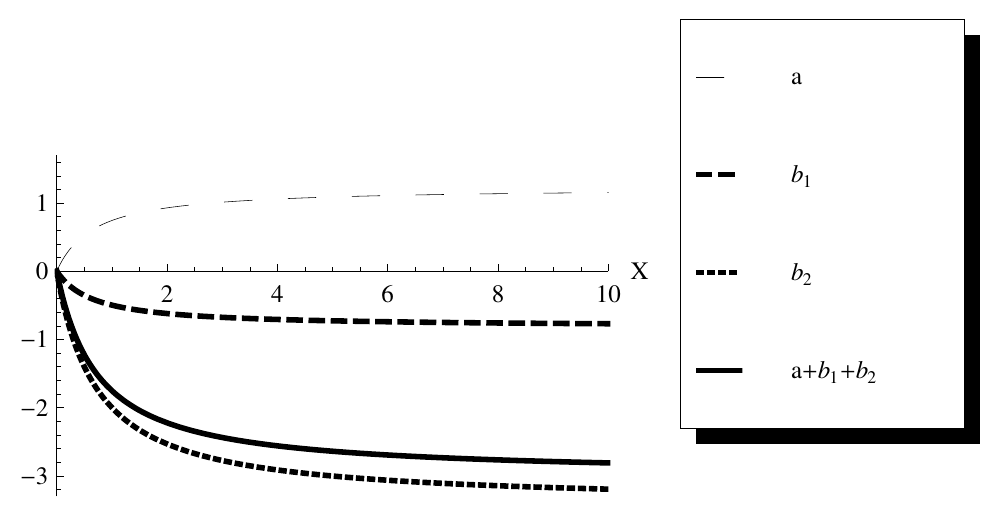}
\caption{The two-loop diagrams $a,\,b_1,\,b_2$ and the two-loop correction $B_1 \, \equiv \ a+b_1+b_2$ as a function of variable $X$ (\ref{eqn_X}). }
\label{2loopplot}
\end{center}
\end{figure}

\section{Summary}
The combined results for the probability to find a particle
at point $x_{0}$ in the quartic double-well potential at zero temperature is
\be
\label{PX0}
    P(x_{0})\sim \frac{e^{-\frac{X^2}{2\la} - \frac{X^3}{3\la}}} {(1+X)^2} \left(1-\la \frac{X (4+3 X) }{(1+X)^2} + O(\la^2)\right) \ ,
\ee
where, we remind, $X=\sqrt{2\la}\, x_{0}$. Note that $X=-1$ is indeed a singularity of the potential
in the unphysical domain.

The $x_0$ dependence of  (\ref{PX0})  is plotted in Fig.
\ref{fig_compare} by the thick line.  The thin line is asymptotics derived in appendix A:
since $x_{0}$-independent constant remained unknown we  normalized it to our curve at
large distances. Their comparison shows good agreement for $x_{0}>1$.

Although derived semiclassically, and thus formally valid for large flucton action only,
our answer is also obviously correct at small $x_{0}$, where it merges with
the answer for harmonic oscillator.

For brevity of the paper we only calculated $P(x_{0})$ at coordinates outside the
two minima. If the barrier is sufficiently large, semiclassical calculation of this probability
can also be extended to the region between the minima. In this case, as noted already
in  \cite{Shuryak:1987tr}, there are four distinct classical trajectories going
through a point: instanton, anti-instanton, and two different fluctons, relaxing
to the left and right minima. Since contributions of those
are additive and have different actions, the probability should be then written as their respective sums.

For completeness we present the corresponding probability $P(x_{0})$ in the case of power-like potential (\ref{V2N}),
{\small
\be
  P(x_{0}) \sim { \exp[-\frac{2\,g\,|x_0|^{N+1}}{N+1} ]  \over |x_0|^N }
  \ ,
\ee}
and the anharmonic oscillator case (\ref{AHpot})
{\small
\be
    P(x_{0})\sim { \exp[- \frac{2}{3} \frac{X_2^3-1}{g} ] \over X_2\,(1+X_2)}  \ ,
    \qquad  X_2=\sqrt{1+g\,x_0^2} \ ,
\ee}
and the sine-Gordon potential
{\small
\be P(x_{0})\sim {\exp[- 16\, \frac{\sin^2(\tilde X)}{g^2} ] \over \exp[ 2\,\tan(\tilde X) ]}\ , \qquad  {\tilde X} = \frac{g\,x_0}{4} \ .
\ee }

\begin{figure}[htp]
\begin{center}
\includegraphics[width=3.0in,angle=0]{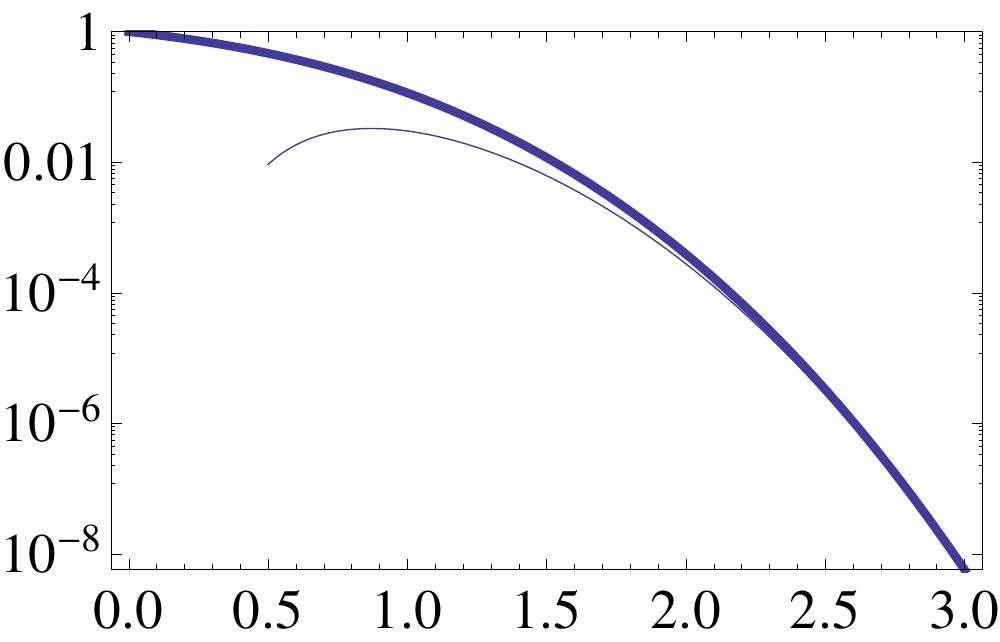}
\caption{The  probability $P(x_{0})$ to find particle at location $x_{0}$ for $\la=0.1$. The thick line
is our result (\ref{PX0}), thin line is asymptotics derived in appendix A. }
\label{fig_compare}
\end{center}
\end{figure}

\section{Discussion} \label{sec_other_apps}

In this section we discuss, without detailed solution, several potential approximation of the developed method.

The simplest extension of what was done is to a QM/SM multidimensional problems is the cases
in which the variables can be separated. For example, multidimensional motion with
a spherically symmetric potential $V(r)$, $r^2=\sum_i x_i^2$, flucton is a classical path along the radial direction, easily calculated from energy conservation (as in 1d case).
However, in this case also one can use WKB or similar approximations as well.

The next problem can be multidimensional anharmonic oscillators, perturbed say by certain cubic and quartic terms. Physical problems of this type are e.g. vibrational states of the multi-atomic molecules. Let us further note that additional appeal of our approach is the fact that in realistic case the temperature $T$ and oscillation quanta $\hbar \omega_i$ are often comparable: so generalization to finite-$T$ fluctons is fully needed. Of course, in this case energy conservation is not enough, and flucton solution should be found from the second-order equations of motion (in Euclidean time) numerically.  This solution should start at a given point of interest $x_{0i}$ and relaxes to the bottom of the potential (the classical vacuum).

Example in which there are additional integrals of motion can be interesting to study as well. For example, an electron in field of two static Coulomb centers has nontrivial integral of motion containing the ``velocities" (first order time derivatives) of coordinates..
Since the motion is in a plane, using the energy and this integral should allow to use the first order equations and perhaps find the flucton solution.

One  would also like to promote the semiclassical method to QFT.
In this case path integral in d-dimensions have a ``boundary value" at $\tau=0$ in form of some $(d-1)$-dimensional field configurations  $\phi(0,\vec{x})$.

To give a simple example, consider scalar field with $\lambda \phi^4/4! $ theory in 4 dimensions. A flucton can be made  of two pieces of Fubini-Lipatov ``instantons" \cite{Fubini:1976jm,Lipatov:1976ny}, shifted away from the $\tau=0$ plane by some equal distance $T/2$
\be
  \phi_{inst}\ =\ 4\sqrt{3\la} {\rho \over x^2+\rho^2 }\ ,
\ee
\be
      \phi_{flucton}(x)= \theta(\tau) \phi_{inst} (\tau+T/2)
\ee
\[
   +\ \theta(-\tau) \phi_{inst} (\tau-T/2)\ .
\]
A configuration at the plane is certain spherically-symmetric ``bump", with a width $\rho$ and a hight $1/\rho$, and the corresponding action is $O(1/\la)$.

The fluctuations around  classical fluctons are described by the quadratic form operator
\be
 O_D \phi \equiv -\partial^m\partial_m{\phi}+ V''(y_{fluct})
 \phi(x) \ ,
\label{OD}
\ee
which includes a Laplacian in all $d$ dimensions. In cases (like the one just mentioned)
when flucton is locally made of instanton solutions, the operator is locally the same, but one should keep in mind that the corresponding Green functions are very different. Indeed, instantons have zero modes and issues related to the orthogonality to those, while in the flucton case the configuration is fixed at $\tau=0$ and thus no zero modes are present. (Therefore, one should not use known instanton Green functions!) 


Finally, let us remind the main flucton idea, applied for any theory.
At $\tau=0$ one may put any $d-1$ dimensional field configuration, and then complement it by $d$-dimensional classical solution, leading to it from the classical vacuum. The exponent of its action provides the probability of the configuration to appear in quantum (or thermal) ensemble. Of course, in many cases, the flucton can be constructed from pieces of known classical solutions, for example out  of pieces of an instanton and an anti-instanton.


In gauge theories the coordinates are gauge fields $A_\mu(x)$, and the $(d-1)$-dimensional field configuration in question should better possess some special properties, which would be gauge independent. Example of a topologically important $(d-1)$-dimensional property can be
the famous  Chern-Simons number $N_{CS}$. Specifically, the so called $sphalerons$ \cite{Klinkhamer:1984di}
are solutions with $N_{CS}=1/2$: there is large literature devoted to the calculations of the probability of its occurrence in quantum or thermal ensembles. We think its calculation by the flucton method would be of interest.

The fluctons should not be confused with other $d$-dimensional paths leading to sphalerons.
In particularly, in \cite{Ostrovsky:2002cg} such paths have been constructed via some
four-dimensional instanton-antiinstanton ``streamline" configurations. This construction is $different$ from fluctons, because the four-dimensional instanton-antiinstanton configuration used satisfies a ``streamline equation" different from classical Yang-Mills equations of motion. It can be called ``forced tunneling", and its objective is to produce $d-1$ configurations with zero kinetic energy, called the ``turning points" configurations, ready to be continued into a Minkowski time.

We hope to be able to address some of these problems in detail in our subsequent publications.

\appendix
\section{Asymptotics at large $x$} \label{sec_asymptotics}

While in the rest of the paper we only apply tools available in QFT settings,
that is path integrals, in this Appendix we still return to the Schr\"odinger equation,
which has the form (here as in the text $\hbar=1, m=1$)
\be
    \left(-\frac{1}{2} \pa_y^2 + V(y)-E \right) \Psi\ =\ 0  \ ,\ \pa_y \equiv \frac{d}{dy}\ ,
\ee
where the double-well potential in shifted coordinates we use is
\be
      V(y) = \frac{y^2}{2} + \sqrt{2\,\la}\, y^3 + \la y^4 \ .
\ee
Note that it smoothly goes to the harmonic oscillator at $\lambda\rightarrow 0$.
Introducing the phase $\phi(y)=-\log \Psi(y)$ we move to the Riccati equation,
\be
     \pa_y^2 \phi - (\pa_y \phi)^2\ =\ 2 E - 2\, V(y)\ ,
\ee
to which one can plug the asymptotic expansion at $|y| \rar \infty$ and obtain all the coefficients (cf. \cite{Turbiner:2010})
\[
     \phi\ =\  \frac{1}{3} \sqrt{2} \sqrt{\la} |y| y^2\ +\  \frac{1}{2}y^2\ -\ d \log|y|^2\ +
\]
\be
  \frac{1 +2 E}{2\,\sqrt{2\,\la}}\ \frac{1}{|y|} - \frac{1}{8\, \la \,  y^2}\ +\ \ldots \ ,
\label{phase}
\ee
where $d=1/2$.
The first two terms in the expansion are classical coming from classical Hamilton-Jacobi equation, log-term reflects an intrinsic property of the Laplacian: $y$ is zero mode or kernel, this term comes from determinant, asymptotically the determinant behaves like $|y|^2$, where $d$ is degree with which it enters to the wavefunction. Note that a constant, $O(x_0^0)$ term is absent: it can not be obtained from the Riccati equation containing derivatives only.
Note also that so far the energy remains undefined: to find it one needs to solve the equation to all $x$. The last terms are true quantum corrections, decreasing at large distances.
Intrinsically, this expansion corresponds to the ground state: it implies that the eigenphase $\phi$ has no logarithmic singularities at real $y$.  Quantization for the Riccati equation implies a search for solutions growing at large $y$ with finite number of logarithmic singularities at real finite $y$. For the $n$th excited state the first two growing terms in
(\ref{phase}) remains unchanged while log-term gets integer coefficient, $(n+1) \log |y|$, see \cite{Turbiner:2010}.

Multiplying by 2 (path integral is for density matrix, or wave function squared) one finds,
as expected, that the first two terms coincide with the classic action of the flucton.
For the determinant one needs to expand at large $x_0$
\be
\label{eqn_aa}
   \log(1+\sqrt{2\la}x_0)\ =
\ee
\[
\log(x_0) + \log(\sqrt{2\la})+ \frac{1}{\sqrt{2\la} x_0 }\ +\ \dots \ ,
\]
and observe that the leading term agrees with the $\log|y|$ term in the asymptotic expansion (\ref{phase}).

The two-loop correction $B_1\,\la$ found in the text (\ref{B1}) expands in inverse powers of $x_0$ as follows
\be
 -\ \la\, \frac{X (4 + 3 X) }{(1 + X)^2}\ =\ -3 \la + \frac{\sqrt{2\la}}{x_0}\ +\ \dots \ ,
\label{eqn_bb}
\ee
where $X = \,\sqrt{2\,\la} x_0$, see (\ref{eqn_X}).

In order to compare the $1/x_0$ terms in the last two expressions one needs to substitute
the ground state energy to $O(\la)$ accuracy
\be
     E\ =\ \frac{1}{2}\ -\  2\la\ +\ \dots \ ,
\ee
to the $O(\frac{1}{x_0})$ term in (\ref{phase}). After that one finds agreement with both $O(\frac{1}{x_0})$  terms given in (\ref{eqn_aa},\ref{eqn_bb}).

Finally, let us add a comment about WKB expression, in which the semiclassical wave function
has in front $1/\sqrt{p(x)}$ where $p$ is momentum. While at large $x$ its leading asymptotics
is correct, as well as that of our determinant, the WKB one has an unphysical singularity at the turning point.  Our determinant, on the other hand, is a smooth function of $x_0$, and it
correctly reproduces the fluctuations till small $x_0$, where it joins with
the harmonic oscillator behaviour.

\section{The sum rule for the instanton Green function \label{sec_sum_rule}}

While the instanton path (\ref{instanton}) depends on both parameters of the quartic potential,
$\lambda$ and $\omega$, in the corresponding fluctuation potential (\ref{inst_V"})
the coupling constant $\lambda$ drops out. As a result, the spectrum and thus the determinant
does not depend on $\lambda$. Therefore, the method we used in the main text to calculate
the determinant via its derivative over $\lambda$ cannot be used.

Nevertheless one can still differentiate the determinant over the remaining parameter $\om$.
As we will show below, it produces a  non-trivial sum rule for the Green function.

The basis for the sum rule is the relation
\[
 \frac{\pa}{\pa \om^2}\log\, Det'\,(O_{instanton})
\]
\be
 =\ -\int d\tau G(\tau,\tau)  \frac{\pa V_{instanton}}{\pa \om^2}\ ,
\label{relation2}
\ee
where $Det'$ stands for determinant with the zero mode excluded. Its value, normalized
to that of the oscillator, is known  \cite{Vainshtein:1981wh}
\be
   \frac{Det'\, O_{instanton}}{Det\, O_{osc}}\ =\ \frac{1}{12 \om^2} \ ,
\ee
and so the l.h.s. of the relation above is $-1/\om^2$. Unfortunately,
this derivative does not depend on the numerical coefficient $1/12$, which we would like to
calculate, so this sum rule is less useful than the one we used for fluctons in the main text.

The derivative of the potential (\ref{inst_V"}) over $\om$ is calculated directly.
The Green's function $G(x,y)$ on top of the instanton solution
\cite{Olejnik,E. Shuryak} is
\begin{widetext}
\be
\label{GF}
             G(x,y) \ =
 G^0(x,y)\bigg[2-xy+\frac{1}{4}|x-y|(11-3xy)
+{(x-y)}^2\bigg]\ +
\frac{3}{8\,\om}(1-x^2)(1-y^2)\bigg[{\log}(2\,G^0(x,y)) -\frac{11}{3}  \bigg] \ ,
\ee
\end{widetext}
expressed in variables
$x \,=\, \tanh(\frac{\om \tau_1}{2}),\,y \,=\,\tanh(\frac{\om\tau_2}{2})\ $,
in which the familiar Green function $G^0= \frac{1}{2\,\om}e^{-\om |t_1-t_2|}$
of the harmonic oscillator looks as follows
\begin{equation}
      G^0(x,y) \ = \ \frac{1}{2\,\om}  \frac{1-|x-y|-x\,y}{1+|x-y|-x\,y} \ .
\label{GF0}
\end{equation}

\begin{figure}[htp]
\begin{center}
\includegraphics[width=3.0in,angle=0]{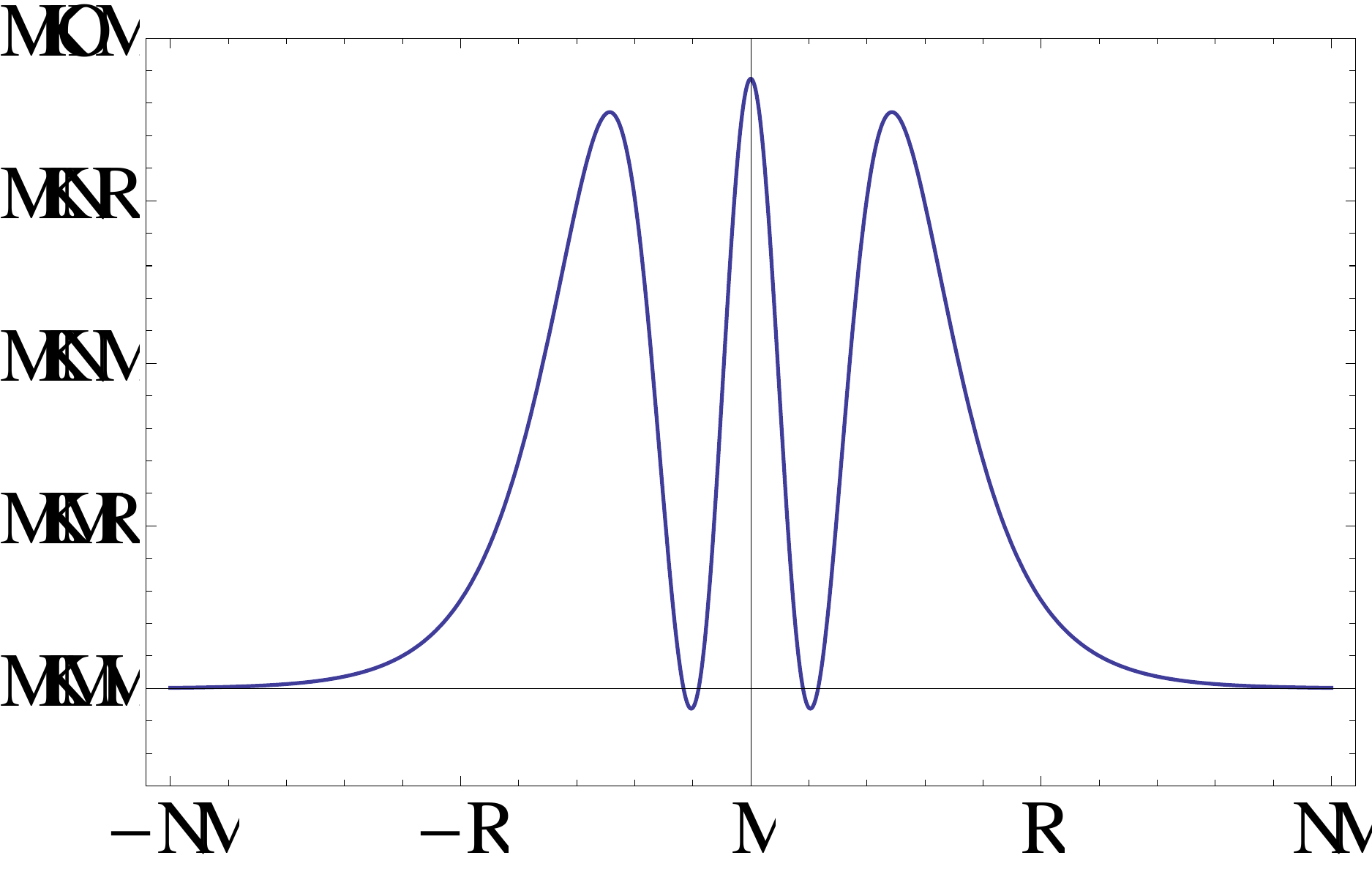}
\caption{The integrand of the sum rule , at $\omega=1$.}
\label{fig_sum_rule}
\end{center}
\end{figure}

We only needs it at the equal arguments $\tau_1=\tau_2$, so it simplifies.
Also one needs to regularize the Green function, by subtracting that of the oscillator,
resulting in
\be
    G_{inst}(\tau,\tau)-G^0(\tau,\tau)={-7 + 4\, \cosh(\tau\, \omega) \over 8\,\omega \,  \cosh^4 (\tau \omega/2)}  \ .
\ee
But even with the simplification, the integrand of the r.h.s. of the sum rule (\ref{relation2}) is rather complicated, see Fig. \ref{fig_sum_rule}. And yet it integrates to unit answer, as the sum rule requires, providing an additional test to the Green function.

\end{document}